\documentclass[conference,9pt]{IEEEtran}
\IEEEoverridecommandlockouts
\usepackage{cite}
\usepackage{amsmath,amssymb,amsfonts}
\usepackage{algorithmic}
\usepackage{graphicx}
\usepackage{textcomp}
\usepackage{xcolor}
\usepackage{multirow}
\usepackage[normalem]{ulem}
\useunder{\uline}{\ul}{}
\usepackage[capitalise]{cleveref}
\def\BibTeX{{\rm B\kern-.05em{\sc i\kern-.025em b}\kern-.08em
    T\kern-.1667em\lower.7ex\hbox{E}\kern-.125emX}}
\ifCLASSOPTIONcompsoc
 \usepackage[caption=false,font=normalsize,labelfont=sf,textfont=sf]{subfig}
\else
 \usepackage[caption=false,font=footnotesize]{subfig}
\fi
\usepackage[export]{adjustbox}

\begin{document}
\bstctlcite{IEEEexample:BSTcontrol}
\title{A Gradient-Interleaved Scheduler for Energy-Efficient Backpropagation for Training Neural Networks\\
\thanks{This research was supported in part by the National Science Foundation under grant number CCF-1814759.}
}

\author{\IEEEauthorblockN{Nanda Unnikrishnan}
\IEEEauthorblockA{\textit{University of Minnesota}\\
Minneapolis, USA \\
unnik005@umn.edu}
\and
\IEEEauthorblockN{Keshab K. Parhi}
\IEEEauthorblockA{\textit{University of Minnesota} \\
Minneapolis, USA\\
parhi@umn.edu}
}

\maketitle
\begin{abstract}
This paper addresses design of accelerators using systolic architectures for training of neural networks using a novel {\em gradient interleaving} approach. Training the neural network involves backpropagation of error and computation of gradients with respect to the activation functions and weights. It is shown that the gradient with respect to the activation function can be computed using a  weight-stationary systolic array while the gradient with respect to the weights can be computed using an output-stationary systolic array. The novelty of the proposed approach lies in interleaving the computations of these two gradients to the same {\em configurable} systolic array. This results in reuse of the variables from one computation to the other and eliminates unnecessary memory accesses. The proposed approach leads to $1.4 - 2.2 \times$ savings in terms of number of cycles and $1.9 \times$ savings in terms of memory accesses. Thus, the proposed accelerator reduces latency and energy consumption.
\end{abstract}

\begin{IEEEkeywords}
Neural Network, Deep learning, Accelerator architectures, Processor scheduling, Gradient interleaving, Systolic array
\end{IEEEkeywords}
\section{Introduction}
Deep neural networks (DNNs) have permeated into all facets of our daily lives as seen in advances for recommender systems, automated photo recognition, and automatic text generations. Inference, in particular, has been extensively investigated because of its inherent advantages in data privacy, response time and bandwidth demand over cloud-based inference. Deep learning networks such as \cite{alexnet,VGG,resnet,googlenet,googleNMT17} have led to a massive surge in data center workloads. Fully connected layers in particular consume over 90\% of the inference workloads currently in  Google's datacenters~\cite{TPUanalysis}.

The common understanding in the community is that machine learning algorithms are efficient because they have a large one time cost and subsequent inference cost is minimal.  This mindset needs to be re-evaluated as we may be severely underestimating the cost of training. There have been numerous architectures that have reported energy consumption for inference \cite{permdnn,eyerissv2_19,EIE} but very few for training \cite{DianNao,DaDianNao14}. The energy consumption for training can be extrapolated from some energy-efficient architectures \cite{eyerissv2_19} as $54 kJ$ for a single epoch considering training a simple network like Alexnet of Imagenet challenge. Thus we can see that even training a small neural network can consume significant energy.

Specifically, it has been shown that compared to the cost of execution, the cost of memory accesses dominates the energy consumption of these accelerators \cite{SzeTutotrial}. This has led to the development of memory-centric schedules that try to minimize the memory bandwidth to the DRAM. This can be achieved by treating it as a caching problem, selecting appropriate block sizes and blocks to maximize the reuse of the SRAM contents \cite{Zhang2d_schedule}. The majority of modern schedulers process the neural network in a layerwise manner and each layer is sequentially scheduled. Further approaches have tried to formulate it as an optimization problem/heuristics by careful partitioning \cite{Part_sced19}. Here, however, optimizations are only limited to intralayer optimizations, which may not exploit the benefits of advantages between layers. This is the case with layer fusion schedules that try to deconstruct the operations within the layer and try to maximize the reuse of variables across layers \cite{layer_fusion19}. Finally, an important aspect is to formulate the dependence graph of the entire training flow and schedule it accordingly. This can maximize parallelism while reducing memory bandwidth \cite{TNPU19}. However, there is a dearth of research that specifically targets the training of the fully-connected layer.

Research on the fully connected layer has been focused on exploiting sparsity during the inference phase \cite{SparseNN18,permdnn}. Unstructured pruning techniques have found some success at the training stage \cite{EagerPrune19}; however, structured sparsity could be better suited for training \cite{permdnn}. Flexible architectures have shown promise at taking advantage of the relative strengths of the different flows at different stages of the CNN \cite{MAERI18, flexflow17, eyerissv2_19}. However, the overall impact is yet to be addressed completely \cite{scalesim2019}. The proposed {\em configurable systolic array} and interleaved scheduler maximize the use of a variable while eliminating intermediate results. The key contribution of this paper is a new scheduling approach for interleaving computations of multiple gradients, as opposed to scheduling these sequentially.

The rest of the paper is structured as follows. \cref{sec:interback} focuses on the backpropagation equation and how systolic arrays and interleaving of gradients can be used to optimize the design.
\cref{sec:eval} evaluates the proposed methodology. Finally, in \cref{sec:conclusion} we summarize the main conclusions of the paper.   

\begin{figure} [!tb]
     \centering
    \includegraphics[width=0.70\linewidth]{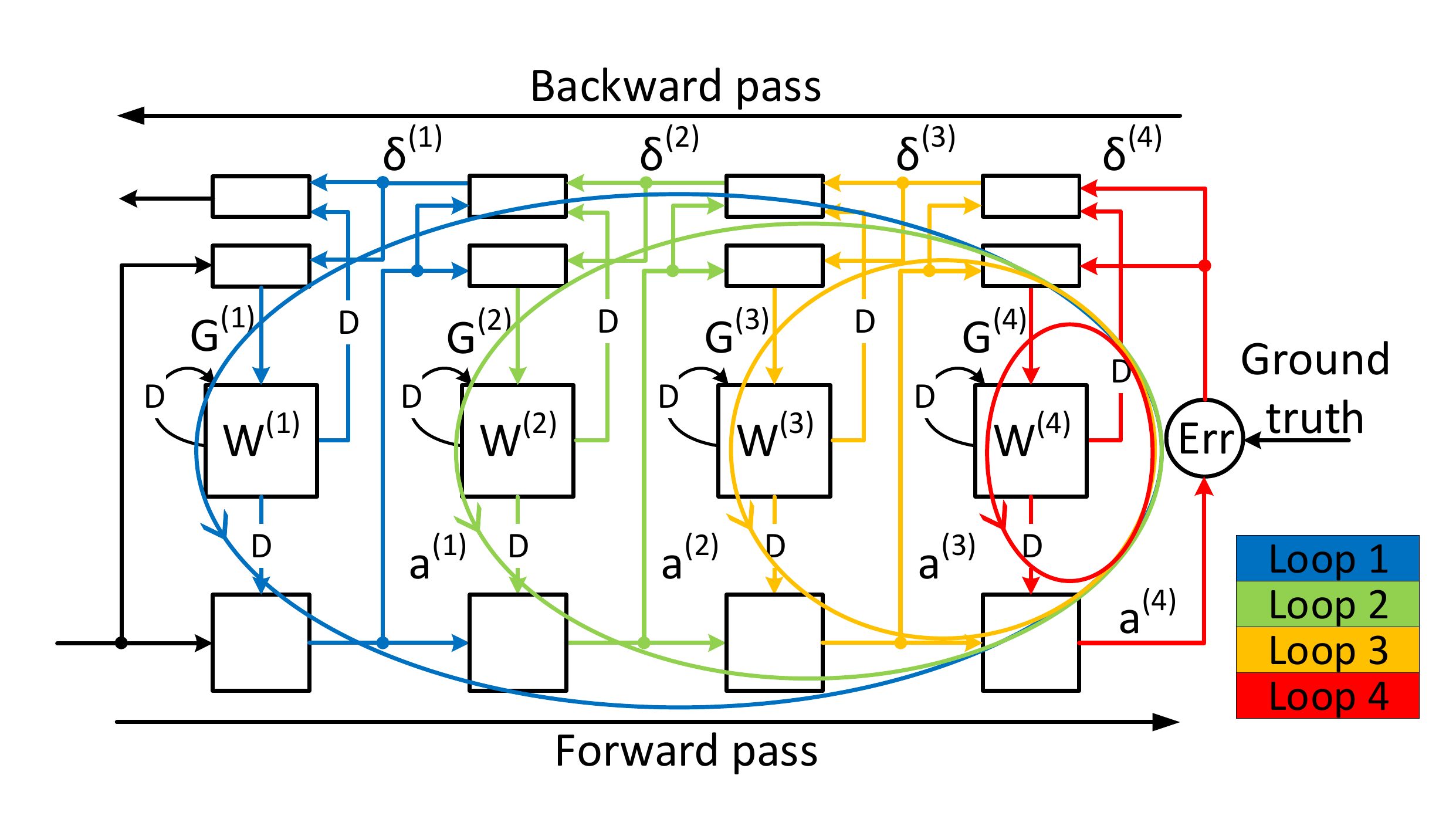}
    \caption{Training loops for a 4-layer fully-connected neural network.}
      \label{fig:layer4_loops}
\end{figure}
\section{Interleaving for backpropagation algorithm}
\label{sec:interback}
The operation of a fully-connected layer can be easily written in terms of a matrix-matrix multiplication. Thus it is natural that many of the traditional scheduling approaches are derived from past implementations. These primarily involve developing blocks or tiles from the matrices and finding the optimal size and order to schedule these blocks. 
Backpropagation involves recursive feedback loops. The iteration period in recursive
computing systems has a fundamental lower bound, referred as the {\em
iteration bound} \cite{iteboundIto95,Parhi99}.

Systolic arrays \cite{kungsys} and other spatial architectures have inherent advantages when it comes to data reuse. Their spatial dataflow pattern allows for more efficient reuse of data. Although current implementations achieve high speedups, these do not exploit all the possible avenues of reuse available. To illustrate this, we focus on the backpropagation algorithm which is the backbone in the training of neural networks. 

The forward pass or inference for the fully-connected layers is well understood and can easily be represented as a matrix-vector multiplication as follows:
\begin{equation}\label{eq:FP}
z^{(l)} = W^{(l)}a^{(l-1)} 
\end{equation}
\begin{equation}\label{eq:FPact}
a^{(l)} = f(z^{(l)})
\end{equation}
where $l$ represents the layer being processed, $W^{(l)}$ is the weight matrix of the fully-connected layer, and $f$ is the activation function for that layer. This paper does not attempt to optimize the forward pass computations and  \cref{eq:FP,eq:FPact} are no longer discussed.

For the fully-connected layer, there are two sets of gradients that need to be computed. The first is to calculate the gradients for each weight matrix. The latter requires computation of the gradient with respect to the activation function. These can be summarized by the following set of equations: 

\begin{equation}\label{eq:gradact}
\delta^{(l-1)} = \frac{\partial E}{\partial a^{(l-1)}} \odot f'(z^{(l-1)})
\end{equation}
\begin{equation}\label{eq:graddelta}
where ~ \frac{\partial E}{\partial a^{(l-1)}} = (W^{(l) T}\delta^{(l)})
\end{equation}
\begin{equation}\label{eq:gradW1}
{G^{(l)}} = \left(\frac{\partial E}{\partial W^{(l)}}\right) = \delta^{(l)} a^{(l-1) T}
\end{equation}
where $E$ represents the loss function, $\delta^{(l)}$ represents the training error gradient backpropagated to layer $l$, and $f'$ represents the derivative of the activation function. The notation $\odot$ represents the Hadamard product of matrices. 

\begin{figure}[!tb]
     \centering
\includegraphics[width=0.75\linewidth]{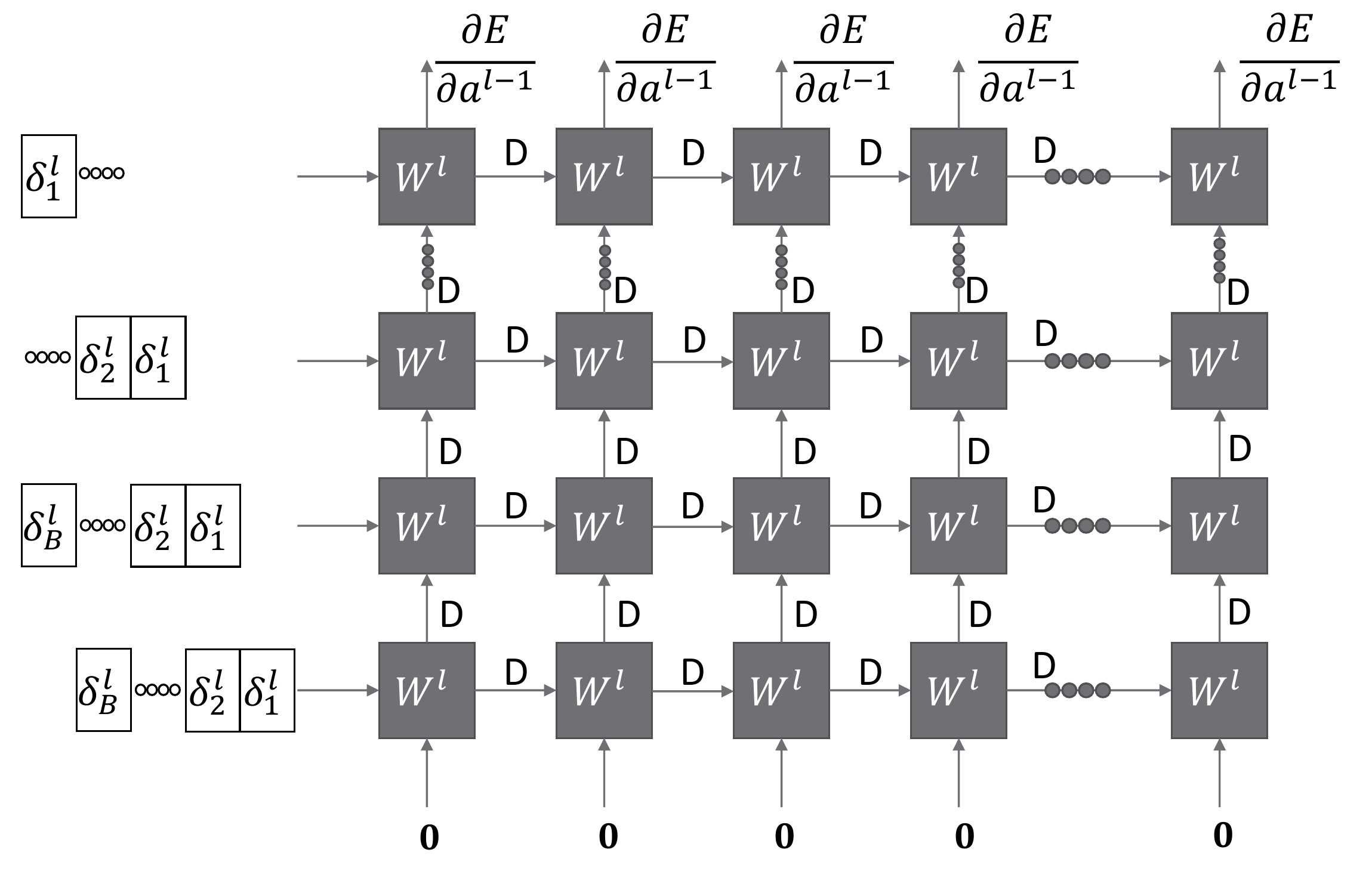}%
\caption{Systolic array that computes $\delta^{(l-1)}$ in a weight-stationary mode.}
\label{fig:graddelta}
\end{figure}

\subsection{Computation of $\delta^{(l-1)}$ }
\label{sec:sysdelta}
To compute \eqref{eq:graddelta} the architecture shown in \cref{fig:graddelta} is considered. The array consists of $P \times Q$ processing elements (PEs), where $P$ and $Q$ represent the horizontal and vertical dimensions of the systolic array. The PEs are interconnected along the horizontal and vertical directions and each contains a pipeline register. The array is set up in a weight-stationary mode, where one of the inputs, $W$, is held constant inside the local memory of the cell. The weights are first loaded into the array from the edge. It takes $Q$ cycles where $P$ words are loaded per cycle into the systolic array. Though in this example only a single weight stored in each processing element is considered, the same technique can be extended to process multiple weights simultaneously. Once the weights are loaded, $\delta^{(l)}$ is loaded at the Western edge, and is staggered by a clock cycle with each row of the design. Once the results are calculated in each PE the partial sums are accumulated vertically in the array. The pipelined architecture is used to assume a simple PE design with a low critical path delay. This proceeds for $B$ more cycles where $B$ is the number of data points in the mini-batch for training. In a traditional design once these calculations are complete the contents of the array are no longer required and are discarded.

\begin{figure}[!tb]
     \centering
\includegraphics[width=0.75\linewidth]{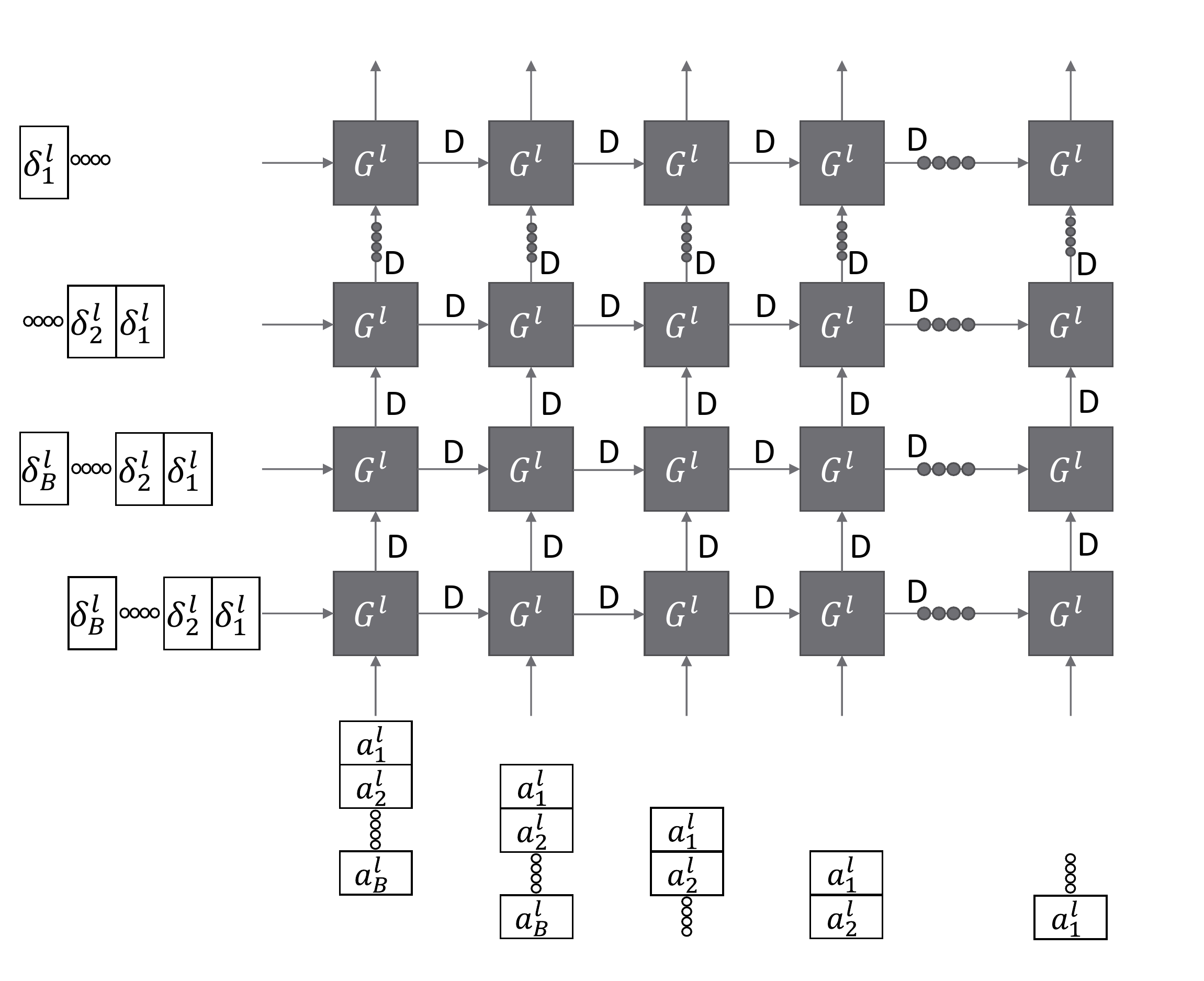}%
\caption{Systolic array that computes $G^{(l)}$ in an output-stationary mode.}
\label{fig:gradW}
\end{figure}

\subsection{Computation of $G^{(l)}$ }
\label{sec:sysG}
To compute \eqref{eq:gradW1} the architecture shown in \cref{fig:gradW} is considered.  To enable a more useful computation, rather than computing \eqref{eq:gradW1} directly, the {\em transpose} of the gradient is computed as follows:

\begin{equation}\label{eq:gradW}
{G^{(l) T}} = \left(\frac{\partial E}{\partial W^{(l)}}\right)^T =  a^{(l-1)} \delta^{(l) T}
\end{equation}

Rewriting \eqref{eq:gradW1} to \eqref{eq:gradW} is the {\em key} to reducing memory access for computing the gradients in the configurable systolic array. Note that most high-level synthesis \cite{HLS_Latte,HLS_HeteroCL,HLS_VTA,HLS_parhi} systems cannot automate this reformulation. As with the earlier case, for \cref{fig:gradW}, the array consists of $P \times Q$ processing elements. The array is set up in an output-stationary mode, where the partial sums $G^{(l)}$ are held constant inside the local memory of the cell. 

\begin{table*}[!tb]
\centering
\caption{Dataflow in the systolic array along the vertical and horizontal directions}
\label{tab:dataflow}
\begin{tabular}{|l|l|l|l|l|l|l|}
\hline
\multicolumn{2}{|l|}{Time}                     & \multirow{2}{*}{$T_0$}           & \multirow{2}{*}{$T_1$}                   & \multirow{2}{*}{$T_2$}                     & \multirow{2}{*}{$T_3$}                     & \multirow{2}{*}{$T_4$}                       \\ \cline{1-2}
Node                          & I/Os           &                               &                                       &                                         &                                         &                                           \\ \hline
\multirow{3}{*}{$PE_{x,y}$}   & $PE_{x-1,y}$   & $\delta_{y,n}$                & $\delta_{y,n}$                        & $\delta_{y,n+1}$                        & $\delta_{y,n+1}$                        & $\delta_{y,n+2}$                          \\ \cline{2-7} 
                              & $PE_{x,y-1}$   & $a_{x,n}$                     & $res_{x,y-1}$                         & $a_{x,n+1}$                             & $res_{x,y-1}$                           & $a_{x,n+2}$                               \\ \cline{2-7} 
                              & $res_{x,y}$    & acc($\delta_{y,n}$$a_{x,n}$) & $res_{x,y-1}$+$\delta_{y,n}$$w_{x,y}$ & acc($\delta_{y,n+1}$  $a_{x,n+1}$)      & $res_{x,y-1}$+$\delta_{y,n+1}$$w_{x,y}$ & acc($\delta_{y,n+2}$$a_{x,n+2}$)          \\ \hline
\multirow{3}{*}{$PE_{x,y+1}$} & $PE_{x-1,y+1}$ & 0                             & $\delta_{y+1,n}$                      & $\delta_{y+1,n}$                        & $\delta_{y+1,n+1}$                      & $\delta_{y+1,n+1}$                        \\ \cline{2-7} 
                              & $PE_{x,y}$     & 0                             & $a_{x,n}$                             & $res_{x,y}$                             & $a_{x,n+1}$                             & $res_{x,y}$                               \\ \cline{2-7} 
                              & $res_{x,y+1}$  & 0                             & acc($\delta_{y+1,n}$$a_{x,n}$)        & $res_{x,y}$+$\delta_{y+1,n}$$w_{x,y}$   & acc($\delta_{y+1,n+1}$  $a_{x,n+1}$)    & $res_{x,y}$+$\delta_{y+1,n+1}$$w_{x,y}$   \\ \hline
\multirow{3}{*}{$PE_{x+1,y}$} & $PE_{x,y}$     & 0                             & $\delta_{y,n}$                        & $\delta_{y,n}$                          & $\delta_{y,n+1}$                        & $\delta_{y,n+1}$                          \\ \cline{2-7} 
                              & $PE_{x+1,y-1}$ & 0                             & $a_{x+1,n}$                           & $res_{x+1,y-1}$                         & $a_{x+1,n+1}$                           & $res_{x+1,y-1}$                           \\ \cline{2-7} 
                              & $res_{x,y+1}$  & 0                             & acc($\delta_{y,n}$$a_{x+1,n}$)       & $res_{x,y-1}$+$\delta_{y,n}$$w_{x+1,y}$ & acc($\delta_{y,n+1}$$a_{x+1,n+1}$)      & $res_{x,y-1}$+$\delta_{y,n+1}$$w_{x+1,y}$ \\ \hline
\end{tabular}
\end{table*}

\begin{figure*}[!tb]
\centering
\subfloat[]{%
\includegraphics[width=0.45\textwidth,valign=t]{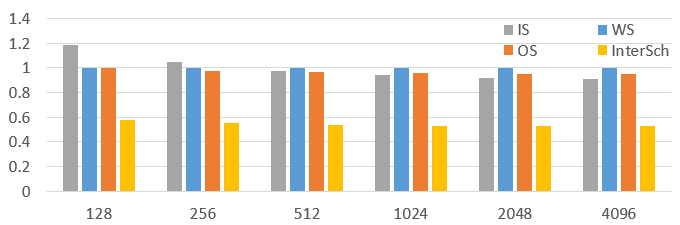}%
}
\hfil
\subfloat[]{\includegraphics[width=0.45\textwidth,valign=t]{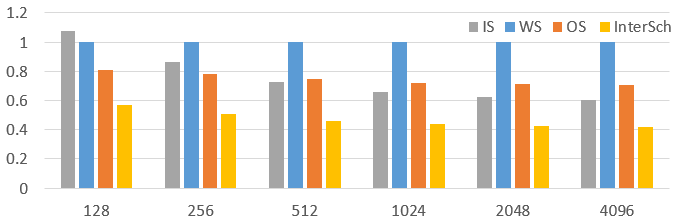}%
}
\caption{Normalized single loop performance of the interleaved scheduler versus traditional dataflow approaches while varying network size (Normalized to WS). a) Normalized number of memory access. b) Normalized number of cycles. InterSch denotes proposed Interleaved Scheduler.}
\label{fig:norm_size}
\end{figure*}
Once again, $\delta^{(l)}$ are passed into the array along the Western edge and the new input $a^{(l-1)}$ is passed into the Southern edge of the array. Both inputs are staggered by a clock cycle and passed into the array and the array processes the inputs in a wavefront manner. The gradients thus generated are accumulated {\em in-place} for the mini-batch of $B$. Once these calculations are complete the contents of the array have to be shifted out. That requires an additional $Q$ cycles where $P$ words are unloaded per cycle from the systolic array.

\subsection{Interleaving of gradients}
\label{sec:sysInter}
To improve performance or reduce the number of memory accesses, traditional approaches have often looked at \eqref{eq:graddelta} to \eqref{eq:gradW} in isolation. However, there are multiple avenues of data reuse to exploit looking at the commonalities between the equations. One way to exploit this is through interleaving, and this has been extensively studied in signal processing systems to reuse hardware and computations across different data points \cite{parhi_fold_synth}. The key approach is to treat the entire backpropagation as a form of a feedback system that can be modeled as a data-flow graph. This enables us to apply the techniques developed in optimizing signal processing systems to improve the backpropagation algorithm. Based on the requirements from \cref{sec:sysdelta,sec:sysG}, this paper proposes to interleave \cite{Parhi99} the gradient computations for \eqref{eq:graddelta} and \eqref{eq:gradW}. To enable this new paradigm of computing we propose to develop {\em configurable systolic arrays} that can switch between different modes of operation on a per cycle basis. This is only possible due to the  reformulation of \eqref{eq:gradW1} to \eqref{eq:gradW} as in both cases $\delta^{(l-1)}$ is input at the Western edge of the array. Thus interleaving \eqref{eq:graddelta} and \eqref{eq:gradW} allows for the reuse of $\delta^{(l-1)}$ across both equations.

\cref{tab:dataflow} summarizes the data flow and operations in the array along the vertical ($y$) and horizontal ($x$) directions. Each node in the array has inputs from the $PE_{x-1,y-1}$, and an output $res$ that represents result. $PE_{x,y}$ and $PE_{x,y+1}$ show the relationship between PEs that are adjacent to one another along the vertical direction. Similarly, $PE_{x,y}$ and $PE_{x+1,y}$ show the relationship between PEs that are adjacent to one another along the horizontal direction. $T_0$, $T_1$ {\em etc}. represent consecutive time steps. From \cref{tab:dataflow} the inherent advantages of the systolic arrays is that once $W$, $\delta^{(l-1)}$ and $a^{(l)}$ are loaded to the array, these are reused multiple times by different PEs. The proposed interleaving also enables $\delta^{(l)}$ to be used in both calculations and it moves with a single cycle delay horizontally. Along the vertical direction the interconnects transfer data alternating between the result for \eqref{eq:graddelta} and \eqref{eq:gradW}. 
This reuse of $\delta$ effectively reduces the number of accesses to the on-chip memory by $B \times \lfloor\frac{N}{Q}\rfloor \times \lfloor{\frac{M}{P}}\rfloor$. Here, $N \times M$ is the dimension of the weight matrix.

Finally, once the gradients are obtained, the weights can be updated as per the following equation:
\begin{equation}\label{eq:gradup}
W^{(l)}(k+1) = F\left(W^{(l)}(k),G^{(l)}(k)\right)
\end{equation}
where $k$ represents the iteration step of the algorithm and $F$ represents the gradient update optimizer such as stochastic gradient descent (SGD), Adam, {\em etc}. From a careful observation of the computations of \eqref{eq:graddelta} and the {\em reformulated} \eqref{eq:gradW} it can be seen that each element of the gradient matrix that is generated from the above architecture is {\em located} in the same PE as the corresponding element of the weight matrix. This is the other advantage of reformulating \eqref{eq:gradW1} as it fortuitously aligns the intended variables. The gradient $G^{(l)}$ is a temporary variable that is generated and must ultimately update the weight matrix; however, due to the conventional approach, it must be stored after creation and recalled from the memory to update the weight as per \eqref{eq:gradup}. However, with the proposed {\em configurable systolic arrays}, the weight matrix can be updated {\em in-place} without the need to store or retrieve the temporary variable $G^{(l)}$. Thus, a further $3 \times B \times \lfloor\frac{N}{P}\rfloor \times \lfloor{\frac{M}{Q}}\rfloor$ accesses are saved.

\begin{figure*} [!tb]
     \centering
    \includegraphics[width=0.70\textwidth]{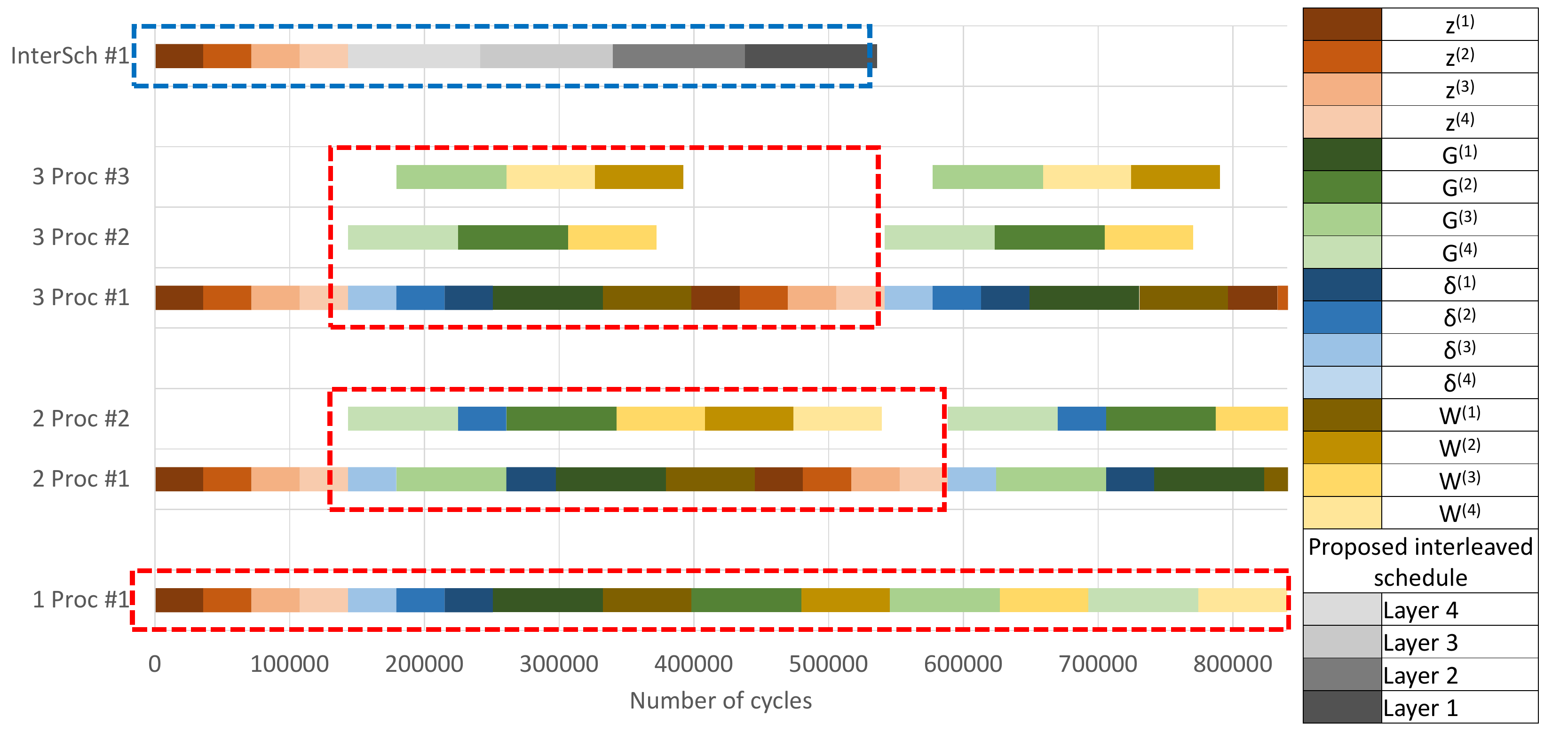}
    \caption{Number of cycles to evaluate the proposed interleaved Scheduler versus baseline output-stationary dataflow for upto 3 Processors.}
      \label{fig:sys_cyc}
\end{figure*}

\section{Performance evaluation of interleaved gradients }
\label{sec:eval}
\subsection{Methodology}
To evaluate the effectiveness of the proposed methodology versus traditional dataflow models, we use the structure shown in~\cref{fig:layer4_loops} as a reference. Each layer in~\cref{fig:layer4_loops} can be modeled for the number of neurons or network size to accommodate various sizes of the fully-connected layer. To enable fast and rapid design space exploration we developed a simulation tool over the open-source python-based NN simulation framework SCALE-sim\footnote{https://github.com/ARM-software/SCALE-Sim} \cite{scalesim2019}.

For evaluation, the underlying PE array size was chosen based on the configuration of Google TPU \cite{TPUanalysis}, a 128x128 systolic array with a matrix-vector multiply unit, and the batch size of the input is configurable. The scope of the proposed work is to limit the number of access to the on-chip memory so all numbers will be reported for the on-chip SRAM without regard to the DRAM access and latency.

\subsection{Single-layer scheduling} \label{sec:SE_SLE}
To measure the single loop efficiency, the innermost loop in~\cref{fig:layer4_loops} (loop 4) is used as a basis for evaluation. For the evaluation of the forward pass~\cref{eq:FP}, as the architecture is flexible, the proposed design is just modeled to match the number of cycles and accesses of the baseline dataflow. For comparison, we evaluate the 3 traditional dataflow models, i.e., weight-stationary (WS), output-stationary (OS), and input-stationary (IS). For the computations in the backward pass, the traditional dataflow computes each equation, \cref{eq:graddelta,eq:gradW,eq:gradup}, as a separate matrix-matrix operation, whereas in the proposed interleaved scheduler methodology the single equivalent time is stated for processing all of the equations in an interleaved manner. Performing \cref{eq:FPact,eq:gradact} are not shown but are assumed to be processed element-wise separately.

\cref{fig:norm_size} analyzes the effect of the network size on the performance of the proposed method. It shows the normalized number of cycles and memory accesses to the local on-chip SRAM. This is obtained by sweeping and evaluating across different network sizes and batch sizes. In \cref{fig:norm_size}, for fixed network sizes, the values obtained are averaged across batch sizes and normalized to the value for weight-stationary. The proposed methodology reduces the overall number of cycles to compute this loop by $30\%$. This corresponds closely to the formulas provide in \cref{sec:sysInter}. Also, the proposed method reduces the number of single-loop memory accesses by $42\%$.


\begin{table}[!tb]
\centering
\caption{Number of cycles and memory accesses for a 4 layer MLP system}
\label{tab:sys}
\begin{tabular}{|r|c|r|c|}
\hline
No of Proc & Cycles ($\times 10^3$) & Utilization     & Memory accesses ($\times 10^6$) \\ \hline\hline
(OS) 1     & 840.08    & 100.00\% & \multirow{3}{*}{129.14}\\
\cline{1-3}
(OS) 2     & 444.96    & 94.12\%  &          \\ 
\cline{1-3}
(OS) 3     & 398.10    & 69.58\%  &          \\ 
\hline \hline
(IS) 1     & 536.00   & 100.00\% &  76.15     \\ \hline
\end{tabular}
\end{table}
\subsection{Multi-layer scheduling}
To measure the system efficiency, all computations for the system in \cref{fig:layer4_loops} are evaluated. At the system level, it is important to consider the dependence graph of all the computations to decide a schedule. Also, in a fully-connected layer, \cref{eq:graddelta,eq:gradW,eq:gradup} can be computed in parallel. Thus to test the design for utilization and parallelizability we investigate 3 scenarios:  a single processor, 2 processors and 3 processors.  In the example shown in~\cref{fig:layer4_loops}, the dependence graph is calculated and a schedule is developed for the above scenarios.

\cref{tab:sys} summarizes the number of memory accesses to the local on-chip SRAM and the number of cycles required to completely process the system with different number of processors for conventional scheduling and using a single processor for the proposed method. As seen in \cref{fig:sys_cyc} and \cref{tab:sys} the proposed method reduces the overall number of processor cycles ((Number~of~processors) $\times$ (cycles/processor)) to compute this loop by $36\%$, $40\%$ and $55\%$ for $1$, $2$ and $3$ processors, respectively. The proposed method reduces the total memory accesses by $41\%$. This corresponds to significant savings in both processor cycles and memory accesses.

\begin{figure} [!tb]
     \centering
    \includegraphics[width=0.9\linewidth]{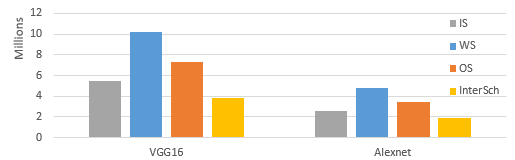}
    \caption{Number of cycles for common CNN architectures.}
      \label{fig:CNN_cyc}
\end{figure}

\begin{figure} [!tb]
     \centering
    \includegraphics[width=0.9\linewidth]{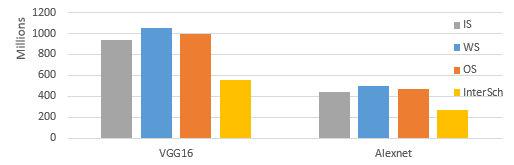}
    \caption{Number of memory accesses for common CNN architectures.}
      \label{fig:CNN_access}
\end{figure}
\subsection{Applications to FC Layers in CNNs}
In order to benchmark the performance of the system we evaluate the proposed method on the fully connected layers of well known convolutional neural networks (CNNs) (VGG16\cite{VGG}, Alexnet\cite{alexnet}). It is shown in \cref{fig:CNN_cyc} that the proposed interleaved scheduler requires $29\%$ less cycles compared to even the best data-flow chosen by a flexible architecture. In terms of memory accesses, the proposed method reduces that requirement by a minimum of $40\%$.

\section{Conclusion}
\label{sec:conclusion}
This paper proposes a novel scheduling scheme based on interleaving the various computations to reduce the latency and memory access of the design. It has been shown that the proposed method outperforms even the best traditional dataflow schemes by a factor of $1.4\times \sim 2.2\times$ in terms of cycles and by a factor of upto $1.9\times$ in terms of memory accesses in fully-connected layers found in common CNNs. Future work will consider the effect of sparsity and specifically structured sparsity on the training process using the proposed approach. Currently, this paper focuses exclusively on feed-forward multi-layer perceptron and it would be of interest to adapt these techniques to the convolutional and recurrent layers as well.

\bibliographystyle{IEEEtran}
\bibliography{references}

\end{document}